\documentstyle[prl,aps,epsfig]{revtex}

\begin{document}

\draft

\title{Fluctuations Indicate Strong Interlayer Coupling in Cuprate Superconductors}

\author{John A. Skinta and Thomas R. Lemberger}
\address{Department of Physics, Ohio State University, Columbus, OH 43210-1106}
\author{E. Wertz, K. Wu, and Q. Li}
\address{Department of Physics, Pennsylvania State University,
University Park, PA 16802}

\date{\today}
\maketitle

\begin{abstract}

From study of the Kosterlitz-Thouless-Berezinskii (KTB) transition
in the superfluid density, $n_s(T)$, of ultrathin \textit{c}-axis
oriented YBa$_{2}$Cu$_{3}$O$_{7-\delta}$ (YBCO) films, we find
that interlayer coupling is unexpectedly strong. The KTB
transition occurs at a high temperature, as if the films were
isotropic rather than quasi-two-dimensional. This result agrees
with a comparison of the superfluid density of YBCO with
Bi$_2$Sr$_2$CaCu$_2$O$_8$ and with numerical simulations of
Josephson junction arrays, and challenges the thermal phase
fluctuation interpretation of critical behavior near $T_c$ in
YBCO.

\end{abstract}


\section{Introduction}

Studies of thermal fluctuations in the phase of the
superconducting order parameter have a long history due to their
importance in understanding phase transitions in general. Because
of the high superfluid densities, $n_s$, and long coherence
lengths, $\xi$, of conventional superconductors, fluctuations in
three-dimensional (3D) samples are significant only within a
hopelessly narrow temperature range near $T_c$ \cite{lobb}, and
most experimental studies involve thin amorphous films in which
reduced dimensionality and superfluid density enhance fluctuations
to an observable level. Two points that are relevant to work on
fluctuations in high-$T_c$ cuprate superconductors have been
established. First, quantum mechanical effects suppress thermal
phase fluctuations (TPF's) when $T$ drops below a surprisingly
high temperature \cite{turneaure01,lemberger,benfatto}. Second, at
high temperatures where fluctuations are classical, simulations of
Josephson junction (JJ) arrays \cite{ohta} are in reasonable
quantitative agreement with measurements of the superfluid density
in two-dimensional (2D) films \cite{turneaure01}. In this paper we
concern ourselves with fluctuations in hole-doped cuprate
superconductors in zero applied magnetic field.

Thermal phase fluctuations were immediately expected to be
significant in cuprate superconductors \cite{lobb}, which have low
superfluid densities and weak interlayer coupling. Exactly how
important TPF's are is a matter of considerable debate. In the
absence of directly comparable experimental results on quasi-2D
conventional superconductors, we rely on simulations of strongly
anisotropic JJ arrays \cite{carlson} to provide a framework for
examining experimental results. In YBa$_{2}$Cu$_{3}$O$_{7-\delta}$
(YBCO), strong fluctuations are evidenced by very rapid changes in
the \textit{ab}-plane penetration depth $\lambda (T)$
\cite{kamal02,anlage,kamal01}, specific heat \cite{charalambous},
and coefficient of thermal expansion \cite{pasler} over a 5 to 10
K interval near $T_c$. Perhaps the most striking evidence that
these phenomena are due to TPF's is power-law behavior with the
3D-XY exponent 2/3 in the superfluid density, $n_s(T) \propto
\lambda^{-2}(T)$, of YBCO crystals \cite{kamal02,anlage,kamal01}.
In the TPF interpretation, fluctuations are strong because of
quasi-two-dimensionality, and interlayer coupling -- although weak
-- begets a crossover from 2D to the observed 3D fluctuations
\cite{carlson,schneider}.

Why not just accept these results at face value and move on? The
reason is that $T_c$ marks a second order phase transition from an
unusual normal state to a {\it d}-wave superconductor. It would
not be surprising if normal-state peculiarities persisted into the
superconducting state and resulted in anomalous $T$-dependence of
$n_s$. For example, if there is a normal-state order parameter
(e.g., see Ref. 13), it may compete with the superconducting order
parameter \cite{rokhsar}.

In the present paper, we argue from measurements of $n_s$ in the
literature and new measurements presented below, that TPF effects
are actually small, presumably because interlayer coupling in
cuprates is much stronger than indicated by transport
measurements.

The pivotal notions of the TPF model are that fluctuations are
strong because interlayer coupling is very weak, and that
interlayer coupling serves only to change the $T$-dependence of
$n_s(T)$ from a discontinuous Kosterlitz-Thouless-Berezinskii
(KTB) \cite{kosterlitz} drop at $T_{KTB}^*$ to a continuous
decrease to zero, with rapidly increasing downward curvature
\cite{carlson,schneider}. One key quantity in comparing $n_s(T)$
with the TPF model is the hypothetical transition temperature,
$T_{KTB}^*$, at which the KTB transition would occur if layers
could be completely decoupled without affecting the pairing
interaction. $T_{KTB}^*$ can be deduced from the measured {\it
ab}-plane superfluid density. A quantitative measure of the
strength of fluctuations, and therefore of interlayer coupling, is
the difference between $T_{KTB}^*$ and the temperature, $T_c$, at
which $n_s(T)$ vanishes: the weaker the coupling, the smaller the
difference. From numerical results in Ref. 6, $T_c /T_{KTB}^* - 1$
increases roughly as $(J_{\perp}/J_{\|})^{1/3}$, where $J_{\perp}$
($J_{\|}$) is the interlayer (intralayer) coupling strength. The
other key quantity is $n_s(T_{KTB}^*) \propto
\lambda^{-2}(T_{KTB}^*)$. In the TPF model, 3D-XY critical
behavior develops only as $n_s$ approaches zero, i.e., for $n_s(T)
\ll n_s(T_{KTB}^*)$.

Let us now take a closer look at power-law behavior in $n_s(T)$
observed for some YBCO samples, setting aside concerns that
power-law behavior is not seen in other YBCO samples
\cite{srikanth,paget} nor in clean Bi$_2$Sr$_2$CaCu$_2$O$_8$
(BSCCO) crystals \cite{lee}. A recent measurement of the {\it
a}-axis magnetic penetration depth in a clean, optimally doped
YBCO crystal \cite{kamal01} finds that at temperatures from $\sim
85$\% to $\sim 99.9$\% of the transition at $T_c \approx 93.78$ K,
$\lambda_a^{-2}(T)$ is fitted to within a percent or so by
$\lambda_a^{-2}(T)/\lambda_a^{-2}(0) \approx 1.26(1 -
T/T_c)^{2/3}$. The critical region encompasses a remarkably large
range of temperature and superfluid density: $0.85 \leq T/T_c \leq
0.999$ and $0.01 \leq \lambda_a^{-2}(T)/\lambda_a^{-2}(0) \leq
0.35$. The argument that an excellent fit is unlikely to be
accidental is taken as support for the TPF model, but in reality
it is not. The reason is that in the TPF model the crossover from
3D to 2D fluctuations occurs well inside the observed ``critical
region''. To see this, let us estimate $T_{KTB}^*$ and
$\lambda_a^{-2}(T_{KTB}^*)$ from the data. With an effective layer
thickness equal to the center-to-center spacing between CuO$_2$
bilayers ($d = 1.17$ nm), $\lambda_a(0) = 160$ nm, and the KTB
prediction \cite{kosterlitz}
\begin{equation}
\label{eq:01} \lambda _{\perp} ^{-1} (T _{KTB}) = \frac{T_{KTB}}{9.8 mm K},
\end{equation}
where $\lambda _{\perp}^{-1} \equiv d\lambda^{-2}$, we find
$T_{KTB}^* \approx 0.96 T_c$ and $\lambda_a^{-2}(T_{KTB}^*)
\approx \lambda_a^{-2}(0)/7$. The TPF model predicts 3D-XY
behavior only for $\lambda^{-2}(T) \ll \lambda^{-2}(T_{KTB}^*)$,
which for these data means $T \geq 0.99 T_c$. Power-law behavior
from 0.85 to 0.99 $T_c$ would have to be ``accidental''. Taking
this one step further, if power-law behavior is due to physics
beyond the TPF model, then TPF's must be so weak as to be
unapparent in the data.

Another argument against the TPF model arises from comparison of
YBCO with BSCCO, which has CuO$_2$ bilayers that are nearly
identical structurally to those of YBCO -- i.e., the same
interplanar spacing and nearly square Cu-O structure -- except
that Y$^{3+}$ is replaced by Ca$^{2+}$ \cite{hazen}. As expected,
the areal superfluid density, $n_s(T)d$, at $T$ = 0 of the
bilayers in clean BSCCO is essentially the same as in the {\it
a}-axis direction (perpendicular to CuO chains) in YBCO. However,
BSCCO has interlayer coupling several hundred times weaker than
YBCO, as measured by transport \cite{iye,chen}, so we expect TPF's
to be much stronger. Let us see. $T_c$ lies 4 K above $T_{KTB}^*$
in YBCO crystals, as found above -- a reasonable result when
compared with simulations of JJ arrays \cite{carlson}. In BSCCO,
the difference is 10 K, indicating that fluctuations in BSCCO are
actually weaker than in YBCO. The latter number comes from clean
BSCCO crystals \cite{lee} in which $\lambda^{-2}(T)$ decreases
linearly with $T$ at low $T$, $T_c$ is 91 K, the center-to-center
spacing between CuO$_2$ bilayers is 1.54 nm \cite{hazen}, and
$\lambda(0) = 210$ nm. Evidently interlayer coupling is stronger
in BSCCO than indicated by transport measurements of {\it
ab}-plane {\it vs}. {\it c}-axis anisotropy.

We could refine the above estimations by using two interlayer
coupling constants, one within bilayers and another between
bilayers, per Carlson {\it et al}. \cite{carlson}, and use an
average of the {\it ab}-plane superfluid density rather than just
the {\it a}-axis value in YBCO, but the conclusions do not change.

Other experimental evidence supports the proposition that
interlayer coupling is stronger than expected. Some corroboration
is provided by a study of $\sigma_1(\omega,T)$ above $T_c$ in YBCO
crystals \cite{anlage}, which finds a critical region of only 0.6
K. More direct support is provided by the measurements presented
below, which demonstrate that fluctuation effects in the 2D
penetration depth, $\lambda _{\perp} ^{-1}(T) \propto n_{S}(T)d$,
of ultrathin twinned YBCO films ($d$ = 4, 8, and 10 unit cells)
are quantitatively consistent with measurements on 2D
\textit{s}-wave films \cite{turneaure01,turneaure03}, assuming
that the YBCO films are effectively isotropic -- i.e., that their
layers are strongly coupled.

\section{Experimental}

The films examined in this study were grown epitaxially by pulsed
laser deposition on NdGaO$_{3}$ substrates, as detailed elsewhere
\cite{kwon}. Each 1 cm $\times$ 1 cm film consists of buffer
layers of semiconducting
Pr$_{0.6}$Y$_{0.4}$Ba$_{2}$Cu$_{3}$O$_{7-\delta}$ that are 12 unit
cells thick above, and 8 unit cells thick below, the YBCO film.
The underlayer lessens the strain of substrate lattice mismatch on
the YBCO, while the capping layer protects the ultrathin YBCO film
from damage during handling. The three films we examined are
nominally 4, 8, and 10 unit cells thick; however, it is likely
that the top and bottom unit-cell layers are not perfectly smooth
or homogeneous, leading to some uncertainty in the film thickness
$d$. We therefore estimate $d = 4\pm1$, $8\pm1$, and $10\pm1$ unit
cells for the three films.

We measure $\lambda _{\perp} ^{-1}(T)$ with a two-coil mutual
inductance technique described in detail elsewhere
\cite{turneaure02}. The films are centered between two
counterwound coils approximately 2 mm in diameter and 1 mm long.
In a typical measurement, the sample is cooled to liquid helium
temperatures, and a current of 100 $\mu$A is driven at 50 kHz
through the coil pressed against the back of the substrate. The
supercurrents created in the sample are very nearly uniform
through the film thickness. Thus, we observe the $T_c$'s of all
layers in the film. As the temperature slowly increases, the
voltage induced in the secondary coil, which is pressed against
the film, is measured continuously. The complex sheet conductivity
$\sigma (\omega ,T)d = \sigma_{1} (\omega ,T)d - i
\sigma_{2}(\omega ,T)d$ of the \emph{entire} film is deduced
directly from the measured mutual inductance. The 2D penetration
depth is obtained with an accuracy of about 3\% from the imaginary
part of the sheet conductivity from $\lambda_{\perp}^{-1}(T)
\equiv \mu _{0} \omega \sigma_{2}(T)d$, where $\mu_{0}$ is the
permeability of vacuum. The uncertainty in film thickness enters
only into calculation of the 3D penetration depth $\lambda ^{-2}
(T)$, and into the 2D penetration depth of \emph{each} unit-cell
layer.

Figure 1 displays $\lambda _{\perp}^{-1}(T)$ as measured at 50 kHz
for each film at optimal doping. An accelerated decrease in
$\lambda _{\perp} ^{-1}$ at high temperatures is most obvious for
the 4 unit cell film, but similar features are apparent very close
to $T_c$ for the 8 and 10 unit-cell films as well. The transitions
are slightly broader than in $\it{a}$-MoGe films
\cite{turneaure01}, presumably due to slight inhomogeneity. The
vertical dotted lines mark the location of the peak in
$\sigma_{1}(50$ kHz,$T)$, whose temperature we define to be $T_c$
and which very nearly coincides with the predicted KTB transition
temperature, $T_{KTB}$, for an isotropic film. To further motivate
association of the drop in $\lambda_{\perp} ^{-1}(T)$ with the KTB
transition, we appeal to the frequency dependences of $\sigma
_{1}$ and $\lambda _{\perp} ^{-1}$ near the transition. Without
going into detail, the observed shift of $\sigma _{1}$ and
$\lambda _{\perp} ^{-1}$ to higher temperature with increasing
frequency, shown in Fig. 2 for the 10 unit cell film, is quite
similar to behavior seen in $\it{a}$-MoGe films
\cite{turneaure03}. The experimental $T_c$ in these ultrathin
films marks a KTB transition.

Arrows in Fig. 1 indicate $T_{KTB}^*$ -- the KTB transition
temperature for uncoupled layers -- showing that $T_{KTB}^*$ lies
8 to 20 K below $T_c$, a difference significantly larger than seen
in YBCO crystals \cite{kamal02,anlage,kamal01} and simulations of
JJ arrays \cite{carlson}. Another indicator of fluctuation effects
is a rapid increase in downward curvature of
$\lambda_{\perp}^{-1}$ for $T > T_{KTB}^*$ as predicted by the TPF
model. The thin solid curves in Fig. 1 are extrapolations of
quadratic fits to the data over $T_{KTB}^* - 10$ K $\leq T \leq
T_{KTB}^*$. We emphasize that the quadratics fit the data
extremely well over at least 20 K below $T_{KTB}^*$, highlighting
the constant curvature of the data below $T_{KTB}^*$. The insets
to Fig. 1 show that only as $T$ gets close to $T_c$ is a change in
curvature of the data apparent. The proximity of the accelerated
drops in $\lambda _{\perp} ^{-1}$ to the predicted $T_{KTB}$, and
the lack of fluctuation effects (e.g., strong downward curvature)
at $T_{KTB}^*$, indicate the presence of strong interlayer
coupling.

We now address an important quantitative issue: the suppression of
$\lambda_{\perp}^{-1}$ below its mean-field value just below $T_
{KTB}$. Taking the thin solid curves in Fig. 1 as reasonable
estimates of the mean-field behavior of $\lambda_{\perp}^{-1}$, we
conclude that $\lambda _{\perp} ^{-1}$ is 35\% to 75\% of its
mean-field value at the transition, in agreement with results from
$\it{a}$-MoGe films \cite{turneaure01} and calculations for 2D JJ
arrays \cite{ohta}.

Thus, the fluctuation behavior of ultrathin YBCO films is
quantitatively similar to that of \textit{a}-MoGe films. Within
the TPF model, the only explanation is that interlayer coupling in
the YBCO films is much stronger than expected from transport
anisotropy. The strong {\it ab}-plane {\it vs}. {\it c}-axis
anisotropy in transport coefficients could be explained
phenomenologically by occasional weak links between adjacent
CuO$_2$ bilayers, with most bilayers coupled strongly to their
neighbors.

\section{Conclusion}

We have presented high precision, low frequency measurements of
$\lambda_{\perp }^{-1} (T)$ and $\sigma_{1}(T)$ in YBCO films of 4
to 10 unit cells thickness. The data indicate that interlayer
coupling is strong, insofar as thermal phase fluctuations are
concerned. Whatever the origin of the coupling, it is the presence
of an additional layer or layers that lessens the effect of
fluctuations on a single unit-cell layer. These results, and
literature results for the \textit{ab}-plane superfluid density of
BSCCO crystals, argue that fluctuations are much weaker in
cuprates than one would expect based on anisotropic transport
coefficients. The power-law behavior near $T_c$ seen in the
superfluid density of very clean YBCO crystals -- cited as
evidence of strong thermal phase fluctuations -- is due to some
other physics, perhaps competition between the superconducting
order parameter and a normal-state order parameter that is
sensitive to disorder. The idea of a normal-state order parameter
has been proposed on various grounds \cite{chakravarty,rokhsar}.
The theory of Rokhsar \cite{rokhsar} suggests that the critical
region may be more complicated than in the usual
superconducting-to-normal transition when the normal and
superconducting order parameters interact.

\section{Acknowledgements}

The authors gratefully acknowledge useful discussions with C.
Jayaprakash and Y.-B. Kim. This work was supported in part by DoE
Grant DE-FG02-90ER45427 through the Midwest Superconductivity
Consortium.


\begin{figure}[htb]
\centerline{\epsfxsize=3.0in \epsfbox{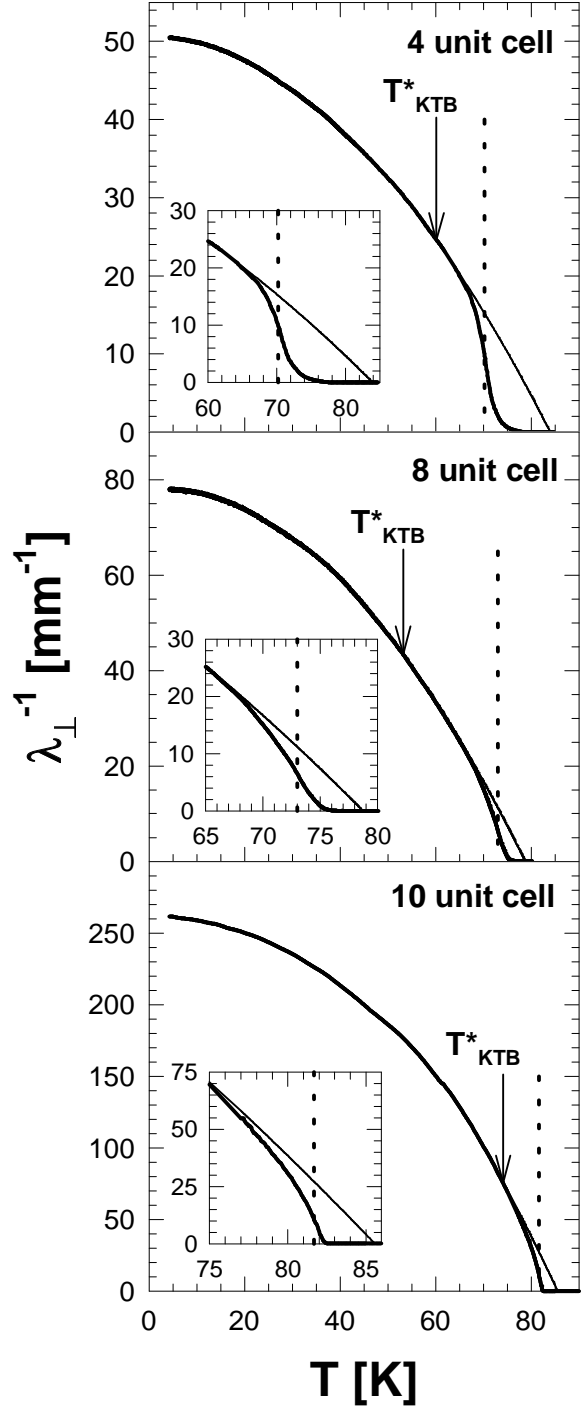}} \vspace{0.1in}
\caption{$\lambda_{\perp}^{-1}$ \textit{vs}. $T$ measured at 50
kHz for three ultrathin YBCO films at optimal doping (thick solid
lines). The vertical dotted lines locate the positions of the
peaks in $\sigma _{1}($50 kHz,$T)$, which very nearly coincide
with $T_{KTB}$. The thin solid lines are extrapolated quadratic
fits to $\lambda_{\perp}^{-1}(T)$ over $T_{KTB}^* - 10$ K $\leq T
\leq T_{KTB}^*$ and provide an estimate of the ``mean-field''
behavior of $\lambda_{\perp}^{-1}$ for $T \geq T_{KTB}^*$. Insets
are enlarged views of the transition regions.} \vspace{-0.1in}
\end{figure}


\begin{figure}[htb]
\centerline{\epsfxsize=3.0in \epsfbox{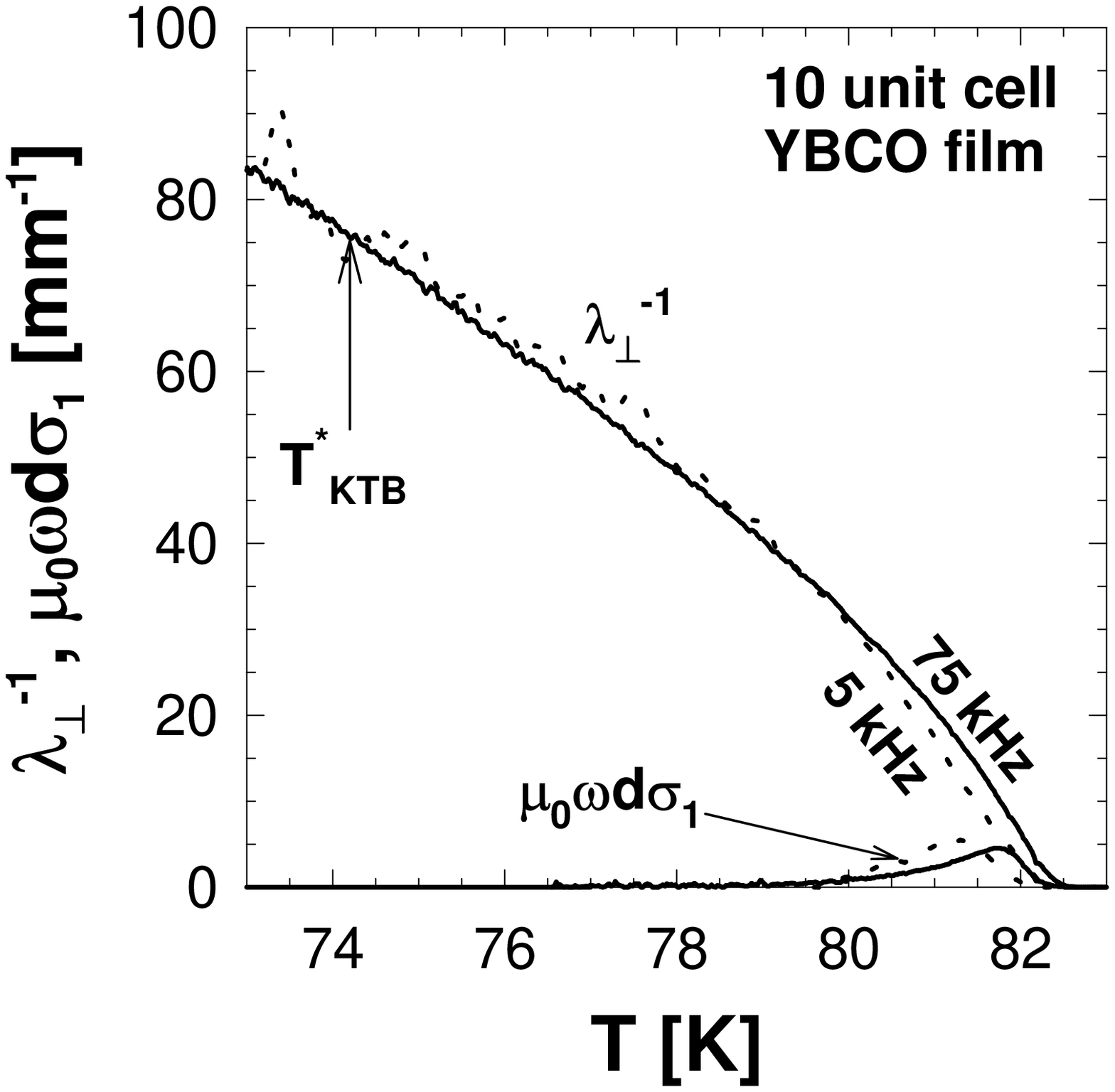}} \vspace{0.1in}
\caption{$T$ dependence of the 2D penetration depth,
$\lambda_{\perp} ^{-1}$, and conductivity, $\mu_{0} \omega d
\sigma _{1}$, at 5 kHz (dotted lines) and 75 kHz (solid lines) for
the 10 unit-cell-thick YBCO film. The arrow indicates
$T_{KTB}^*$.} \vspace{-0.1in}
\end{figure}


\begin{references}

\bibitem{lobb} C.J. Lobb, Phys. Rev. B {\bf 36}, 3930 (1987).
\bibitem{turneaure01} S.J. Turneaure, T.R. Lemberger, and J.M.
Graybeal, Phys. Rev. Lett. {\bf 84}, 987 (2000).
\bibitem{lemberger} T.R. Lemberger, A.A. Pesetski, and S.J.
Turneaure, Phys. Rev. B {\bf 61}, 1483 (2000).
\bibitem{benfatto} L. Benfatto, S. Caprara, C. Castellani, A. Paramekanti,
and M. Randeria, Phys. Rev. B 63, 174513 (2001).
\bibitem{ohta} T. Ohta and D. Jasnow, Phys. Rev. B {\bf 20}, 139
(1979); S. Teitel and C. Jayaprakash, Phys. Rev. Lett. {\bf 51},
1999 (1983); Phys. Rev. B {\bf 27}, 598 (1983); W.Y. Shih and D.
Stroud, {\it ibid.} {\bf 32}, 158 (1985); I.-J. Hwang and D.
Stroud, {\it ibid.} {\bf 57}, 6036 (1998).
\bibitem{carlson} E.W. Carlson, S.A. Kivelson, V.J. Emery, and E. Manousakis,
Phys. Rev. Lett. {\bf 83}, 612 (1999).
\bibitem{kamal02} S. Kamal, D.A. Bonn, N. Goldenfeld, P.J. Hirschfeld, R.
Liang, and W.N. Hardy, Phys. Rev. Lett. {\bf 73}, 1845 (1994).
\bibitem{anlage} S.M. Anlage, J. Mao, J.C. Booth, D.H. Wu, and J.L. Peng,
Phys. Rev. B {\bf 53}, 2792 (1996).
\bibitem{kamal01} S. Kamal, R. Liang, A. Hosseini, D.A. Bonn, and W.N. Hardy,
Phys. Rev. B {\bf 58}, R8933 (1998).
\bibitem{charalambous} M. Charalambous, O. Riou, P. Gandit, B. Billon, P.
Lejay, J. Chaussy, W.N. Hardy, D.A. Bonn, and R. Liang, Phys. Rev. Lett.
{\bf 83}, 2042 (1999).
\bibitem{pasler} V. Pasler, P. Schweiss, C. Meingast, B. Obst, H. W\"uhl,
A.I. Rykov, and S. Tajima, Phys. Rev. Lett. {\bf 81}, 1094 (1998).
\bibitem{schneider} T. Schneider and J.M. Singer, Physica
(Amsterdam) {\bf 341-348C}, 87 (2000).
\bibitem{chakravarty} S. Chakravarty, R.B. Laughlin, D.K. Morr, and C. Nayak,
Phys. Rev. B {\bf 63}, 094503 (2001).
\bibitem{rokhsar} D.S. Rokhsar, Phys. Rev. Lett. {\bf 70}, 493 (1993).
\bibitem{kosterlitz} J.M. Kosterlitz and D.J. Thouless, J.
Phys. C {\bf 6}, 1181 (1973); J.M. Kosterlitz, {\it ibid} {\bf 7},
1046 (1974); V.L. Berezinskii, Sov. Phys. JETP {\bf 32}, 493
(1971).
\bibitem{srikanth} H. Srikanth, Z. Zhai, S. Sridhar, A. Erb, and E. Walker,
Phys. Rev. B {\bf 57}, 7986 (1998).
\bibitem{paget} K.M. Paget, B.R. Boyce, and T.R. Lemberger, Phys.
Rev. B {\bf 59}, 6545 (1999).
\bibitem{lee} S.-F. Lee, D.C. Morgan, R.J. Ormeno, D.M. Broun, R.A. Doyle,
J.R. Waldram, and K. Kadowaki, Phys. Rev. Lett. {\bf 77}, 735
(1996).
\bibitem{hazen} R.M. Hazen, in {\it Physical Properties of High
Temperature Superconductors II}, edited by D.M. Ginsberg (World
Scientific, Singapore, 1990).
\bibitem{iye} Y. Iye, in {\it Physical Properties of High Temperature
Superconductors III}, edited by D.M. Ginsberg (World Scientific,
Singapore, 1992).
\bibitem{chen} X.H. Chen, M. Yu, K.Q. Ruan, S.Y. Li, Z. Gui, G.C. Zhang,
and L.Z. Cao, Phys. Rev. B {\bf 58}, 14219 (1998).
\bibitem{turneaure03} S.J. Turneaure, T.R. Lemberger, and J.M.
Graybeal, Phys. Rev. B {\bf 63}, 174505 (2001).
\bibitem{kwon} C. Kwon, Q. Li, X.X. Xi {\it et al}., Appl. Phys. Lett.
{\bf 62}, 1289 (1993); C. Kwon, Q. Li, I. Takeuchi {\it et al}.,
Physica (Amsterdam) {\bf 266C}, 75 (1996).
\bibitem{turneaure02} S.J. Turneaure, E.R. Ulm, and T.R.
Lemberger, J. Appl. Phys. {\bf 79}, 4221 (1996); S.J. Turneaure,
A.A. Pesetski, and T.R. Lemberger, {\it ibid.} {\bf 83}, 4334
(1998).

\end{references}
\end{document}